 \documentclass[final,5p,times,twocolumn]{elsarticle}

\usepackage{graphicx}
\usepackage{amssymb}
\usepackage{xcolor}

\newcommand{\fixmetext}[1]{{\color{black}#1}}

\journal{Journal of Information Security and Applications}

\begin{document}

\begin{frontmatter}


\title{The Language of Biometrics: Analysing Public Perceptions}



\author[1]{Oliver Buckley}
\author[2]{Jason R. C. Nurse}

\address[1]{School of Computing Science, University of East Anglia, Norwich, NR4 7TJ, UK}
\address[2]{School of Computing, University of Kent, Canterbury, CT2 7NF, UK}

\begin{abstract}
There is an increasing shift in technology towards biometric solutions, but one of the biggest barriers to widespread use is the acceptance by the users. In this paper we investigate the understanding, awareness and acceptance of biometrics by the general public. The primary research method was a survey, which had 282 respondents, designed to gauge public opinion around biometrics. Additionally, qualitative data was captured in the form of the participants' definition of the term \textit{biometrics}. We applied thematic analysis as well as an automated Word Vector analysis to this data to provide a deeper insight into the perceptions and understanding of the term. Our results demonstrate that while there is generally a reasonable level of understanding of what biometrics are, this is typically limited to the techniques that are most familiar to participants (e.g., fingerprints or facial recognition). Most notably individuals' awareness overlooks emerging areas such as behavioural biometrics (e.g., gait). This was also apparent when we compared participants' views to definitions provided by official, published sources (e.g., ISO, NIST, OED, DHS). Overall, this article provides unique insight into the perceptions and understanding of biometrics as well as areas where users may lack knowledge on biometric applications.\end{abstract}

\begin{keyword}
Biometrics \sep thematic analysis \sep word vector analysis \sep security \sep privacy \sep perceptions \sep usable security \sep user study


\end{keyword}

\end{frontmatter}


\section{Introduction}
\label{introduction}

Biometric technologies are becoming increasingly commonplace in our everyday lives across a wide range of applications from securing our personal devices through to managing physical access. For example, a study by Techpinions \cite{techpinions16} revealed that Apple's Touch ID fingerprint technology is used by 89\% of users, with a Touch ID capable device. More recent research by Deloitte discovered that 79\% of UK smartphone owners in general (i.e., iPhone and others) use their device's fingerprint scanner, and more than a third of smartphones now have a fingerprint reader \cite{deliotte2017}. 

It is entirely possible that users are now being exposed to biometric technologies without ever realising it. Since 2016, Barclays Bank has used voice recognition software for its personal telephone banking customers, and other banks such as HSBC have also introduced the technology \cite{bbcBarclaysVoice2016} \cite{tele17}. 

As our reliance on the Internet increases as does the requirement to create memorable, yet hard to guess security credentials. Research conducted by Grawmeyer and Johnson \cite{grawemeyer2011} in 2011 found that an average user will have to manage 7.95 passwords. In contrast to this, a survey carried out in 2014 as part of Cyber Streetwise (now known as Cyber Aware \cite{cyberaware}), a UK government initiative designed to drive behavioural change in cyber security, found that an average user will manage 19 different passwords. This highlights a significant growth in the number of credentials a user is required to manage, with the number of accounts more than doubling in three years. Additionally, there is also the added overhead of authenticating to a wide range of services. Wash et al. \cite{wash2016} found that on average a user will typically be confronted with a password event anywhere between 8 and 23 times per day. 

According to Gehringer \cite{gehringer2002}, it is notoriously difficult to create memorable yet secure passwords. A strong password can be effective but it is all too common for this protection to be compromised by the users themselves. For example, a study by SplahData \cite{SplashData} reveals that `123456' and `password' are still amongst some of the most commonly used passwords. The more traditional methods of security will typically focus on something the user knows (e.g. a username and password) or on something the user possesses (e.g., an RSA token \cite{RSA}). Conversely, the use of biometrics can potentially alleviate many of the issues surrounding security credentials, as the emphasis is placed on what the user is, rather than what the user knows or possesses. When compared to  more traditional approaches to authentication, biometrics claim to offer a greater level of confidence that the individual is who their credentials claim them to be. Additionally, according to Jain et al. \cite{jain2006} biometrics can potentially prove to be more reliable than standard authentication methods and do not rely on credentials or tokens that can be inadvertently lost or stolen.

With our ever-increasing reliance on digital services and the ubiquity and affordability of biometric technologies then it would seem that the solution is obvious. The use of these biometric technologies should make it easy to deliver user authentication that is both secure and cost effective, while at the same time reducing the cognitive load on the users of these systems and services. A crucial part of any increased use of biometrics however, is their acceptance by the general public. 

Despite Apple's impressive statistics on user uptake there are still a number of common misconceptions about the way in which biometric authentication actually works, as discussed by both Ashbourn \cite{ashbourn2014} and Thompson et al. \cite{thompson2018student}. For example, concerns about biometric data being lost or compromised are commonplace, with users often assuming that fingerprints are stored exactly within their phone as an image. However, the reality is that when collecting a fingerprint the data is encoded in some way based on a number of extracted features. These fundamental misconceptions about this technology, particularly when it is so widely available, suggest that despite the increased prevalence of biometric technologies that they are still not widely understood.

In this paper therefore, we engage with members of the general public to investigate their understanding, awareness and acceptance of biometrics. Our goal is to provide useful insight into these factors as well as key related technologies. The remainder of this article is structured as follows. In Section \ref{lit_review} we provide a discussion of the related material, including previous studies of user perceptions. Section \ref{method} provides an overview of the methodology used to conduct the study. Section \ref{analysis_and_discussion} presents the results and discusses their significance, and finally Section \ref{conclusions_and_future_work} concludes this research.

\section{Related Work}
\label{lit_review}
Furnell and Evangelatos \cite{furnell2007} present a survey that explored the awareness and perceptions of biometrics. The survey looked to understand which techniques were commonly understood, which had been used and how they could be used in practice. This work highlighted that the respondents (of which they were 209) were generally positive about the notion of biometrics but with relatively limited practical experience of using them. Additionally, they found that there were a number of technologies that were generally better accepted than others. For example, participants were general more comfortable with the use of fingerprints than retinal scans. While it is over a decade since this seminal work, it will provide an interesting comparison with the research presented in our current article. As previously discussed, there is a greater exposure to biometrics in our daily lives and as such it is anticipated that the acceptance of biometrics will have changed accordingly. 

Work presented by Chan and Elliot \cite{chan2016} provides an updated look at the privacy perceptions of biometrics using two surveys. The first survey, with 200 participants, asked respondents about their experiences and perceptions of biometrics. A second survey, carried out a year later, looked to measure any changes in perceptions over the course of time. This research also suggests a level of skepticism around the security and privacy of their biometric data. For example, over 45\% of respondents would not trust their data with a public corporation. One of the key findings in this research was that there was a greater support for the use of biometrics in both counter-terrorism and banking. 

The work of Sabharwal \cite{sabharwal2016} focuses explicitly on the perceptions of biometrics in banking customers. This research used a survey to understand the concerns, opinions and perceptions of banking customers with respect to e-banking. The results from the survey suggest a number of key metrics when considering the large-scale deployment of biometric technology within e-banking including: reliability, performance, resistance to circumvention and privacy issues.  

One of the areas where there has been a significant increase in the ubiquity of biometrics is in our own personal devices, with fingerprint recognition being very much a standard feature on a modern smartphone or tablet. Bhagavatula et al. \cite{bhagavatula2015} conducted a lab study and survey to determine the usability of smartphone specific biometrics (e.g., fingerprint and facial recognition). Their work highlighted that the majority of users actually preferred the use of fingerprint unlock over facial recognition or the use of a PIN. Similarly, it was found that users perceived fingerprint unlocking to be more secure and convenient than the use of a PIN.  

Research by Krol et al. \cite{krol2016} focuses specifically on the acceptability of face biometrics as a replacement for CAPTCHAs \cite{captcha}. The work used a lab study to test a range of human verification mechanisms, and then used surveys and interviews to determine the acceptance and perceptions of the various techniques. One of the key findings of this work was that the users were generally concerned about the use of their own personal image in verification, highlighting that the privacy of personal data is a key concern to potential users of biometric identification.

Ogbanufe and Kim \cite{ogbanufe2018comparing} focus on user perceptions of the differences between biometric and more traditional methods of authentication for e-payments. Their work found that the biometrics authentication method significantly influenced the security concerns of an individual as well as the perceived usefulness and trust of the online store. It is interesting to note that their research found that users considered fingerprints to be more secure than a combination of credit card and PIN. 

The use of biometrics is not limited to a user's personal life and these are technologies that are becoming increasingly common within the workplace. Carpenter et al. \cite{carpenter2018privacy} presents a study focusing on the privacy concerns of employees related to organisational use of biometrics. Their results highlighted that the self-construal (the extent to which the self is defined independently of others or interdependently with others \cite{cross2011and}) played a significant role in the formulation of privacy, perceived accountability and perceived vulnerability concerns. Their work suggested that they also act as notable indicators of the user's attitude towards biometric technologies in the workplace. 

One of the common themes that is notable across the studies is that users have concerns about the privacy and security of their own personal data. This is something that is further investigated in the research presented in this paper as it explores the concerns of the participants and the contextual nature of these concerns. For instance, this work investigates whether the situation determines the level of security that a user feels is required. The current research to date, in the related literature, has been mainly focused on the perceptions of usability, security and privacy of biometric technologies. The work presented here aims to draw out the public perceptions of biometric technologies and how these perceptions are linked to acceptance. Previous work in this area has not attempted to ascertain users understanding of the term \textit{biometrics}.  In the study presented here, we have built upon a traditional survey with the use of thematic analysis \cite{braun2006using} and a vector representation of words \cite{mikolov2013efficient} to better understand the term \textit{biometrics} and its implications for users. 

\section{Methodology}
\label{method}
To structure our research, we adopted a methodology consisting of common data gathering and analysis processes. Prior to our study commencing it was also reviewed by our university's ethical review board. For recruitment, we used a mixture of convenience and snowball sampling \cite{goodman1961} to gather participants from the general public. This involved advertisements on social media including Twitter, LinkedIn and Reddit. Surveys were selected to allow the gathering of data; the primary reason for this being their ease of deployment and larger participant reach. We designed our survey such that questions first covered a range of demographics including: Age, Gender, Highest level of education, and Area of work or study. On the topic of biometrics, we asked participants whether they had some understanding of the term, and if they did, they were asked to provide a definition.

The survey contained a further eight questions that focused on the participant's awareness and perceptions of various biometric technologies. We asked participants to rank five generic situations \fixmetext{(or usage scenarios)}, chosen to provide different data contexts, based on their requirement for `security'. \fixmetext{Specifically, the situations were Banking, Online shopping, Airport, Mobile device, Home. The choice of these scenarios was arbitrary and primarily motivated by having a set of scenarios that could be ordered by their perceived need for security. We view a potential ordering of these scenarios as follows from most to least: Banking, Airport, Home, Mobile Device, online shopping.} Participants were then queried about which biometrics they would be comfortable using in each of the five situations. This was used to develop an understanding of the technologies that the participants viewed as the most secure and those that they viewed as the least secure. The survey next focused on the perceptions of the security of biometrics when compared to other common techniques, such as passwords and two-factor authentication.  

Our approach to data analysis involved a combination of quantitative and qualitative techniques. \fixmetext{First we applied statistical methods to analyse responses to close-ended questions. Tests for correlation across some aspects (e.g., between education level and familiarity with various methods) were also conducted using} the Pearson Chi-Squared test. 

Next, \fixmetext{the definitions of biometrics provided by each participant were analysed using both manual and automated qualitative techniques. Thematic analysis is a manual data analysis technique, which focuses on allowing the assessment of qualitative data for common themes and patterns~\cite{braun2006using}. This a well-known technique that has been applied across a variety of fields. In addition to this approach we also made use of an automated analysis method, to provide a comparison. We used an approach called Word2Vec~\cite{mikolov2013efficient}; this technique models words in a vector space, allowing for additional insight into textual content.}

\section{Findings and Discussion}
\label{analysis_and_discussion}
\subsection{Survey Results}
\label{subsec:survey_results}
There were a total of 282 participants in our study,  which is in line with other similar studies discussed in Section~\ref{lit_review}. \fixmetext{As part of our analysis we used the Pearson Chi-Squared test to establish if any correlations were present in the data, however, we found there were no significant correlations of note.} 

\begin{figure}[ht!]
\centering
    \includegraphics[width=0.85\linewidth]{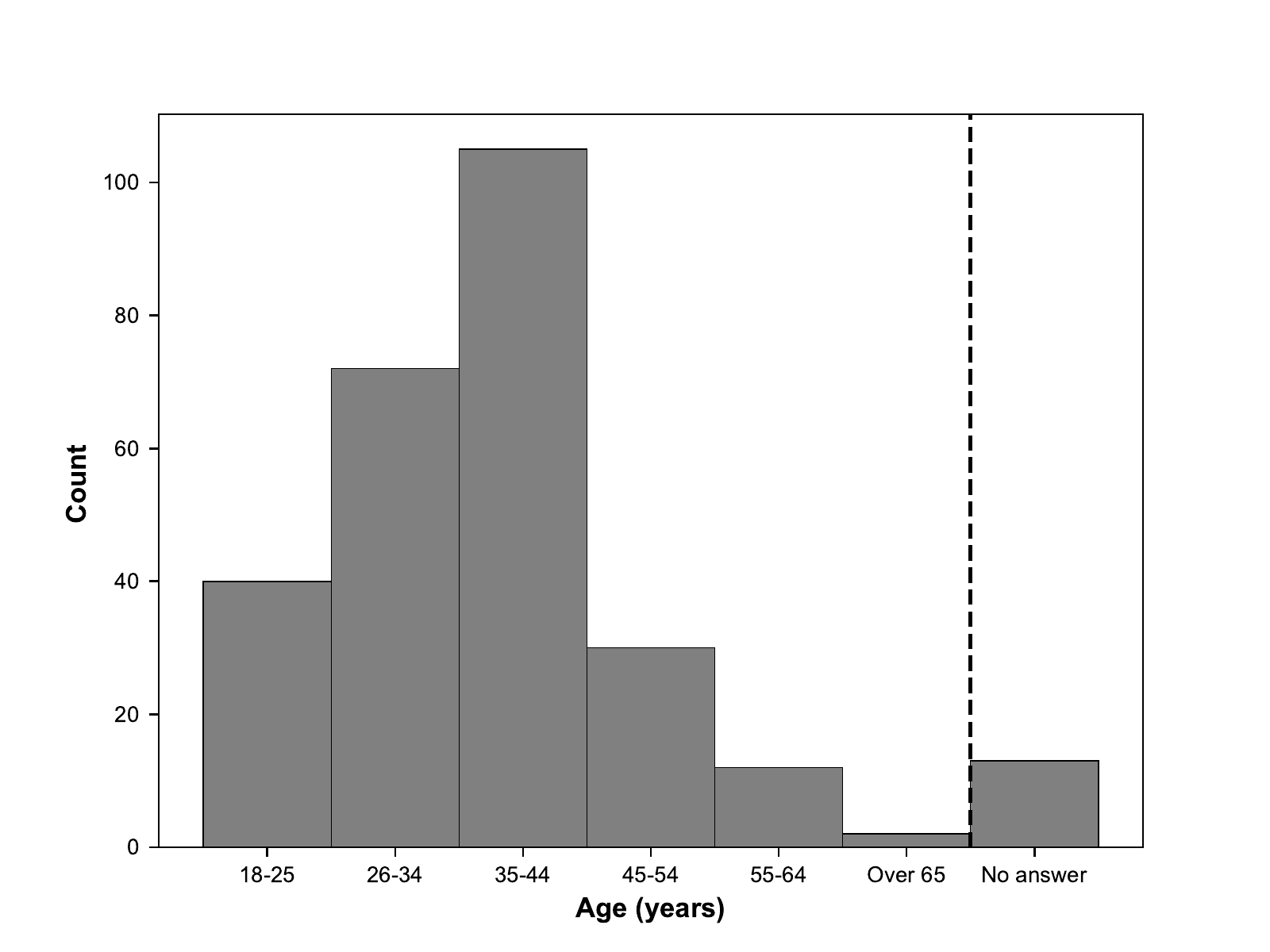}
    \caption{Age of respondents}
    \label{fig:age} 
\end{figure}

The majority of respondents to the survey were aged between 35 and 44 (40\%), as shown in Figure \ref{fig:age}, with approximately 83\% of all participants being aged under 45. However, this distribution of the participants' ages means that the vast majority of respondents will have either grown up with technology from an early age or been early adopters of new technologies. 

\begin{figure}[ht!]
\centering
  \includegraphics[width=0.85\linewidth]{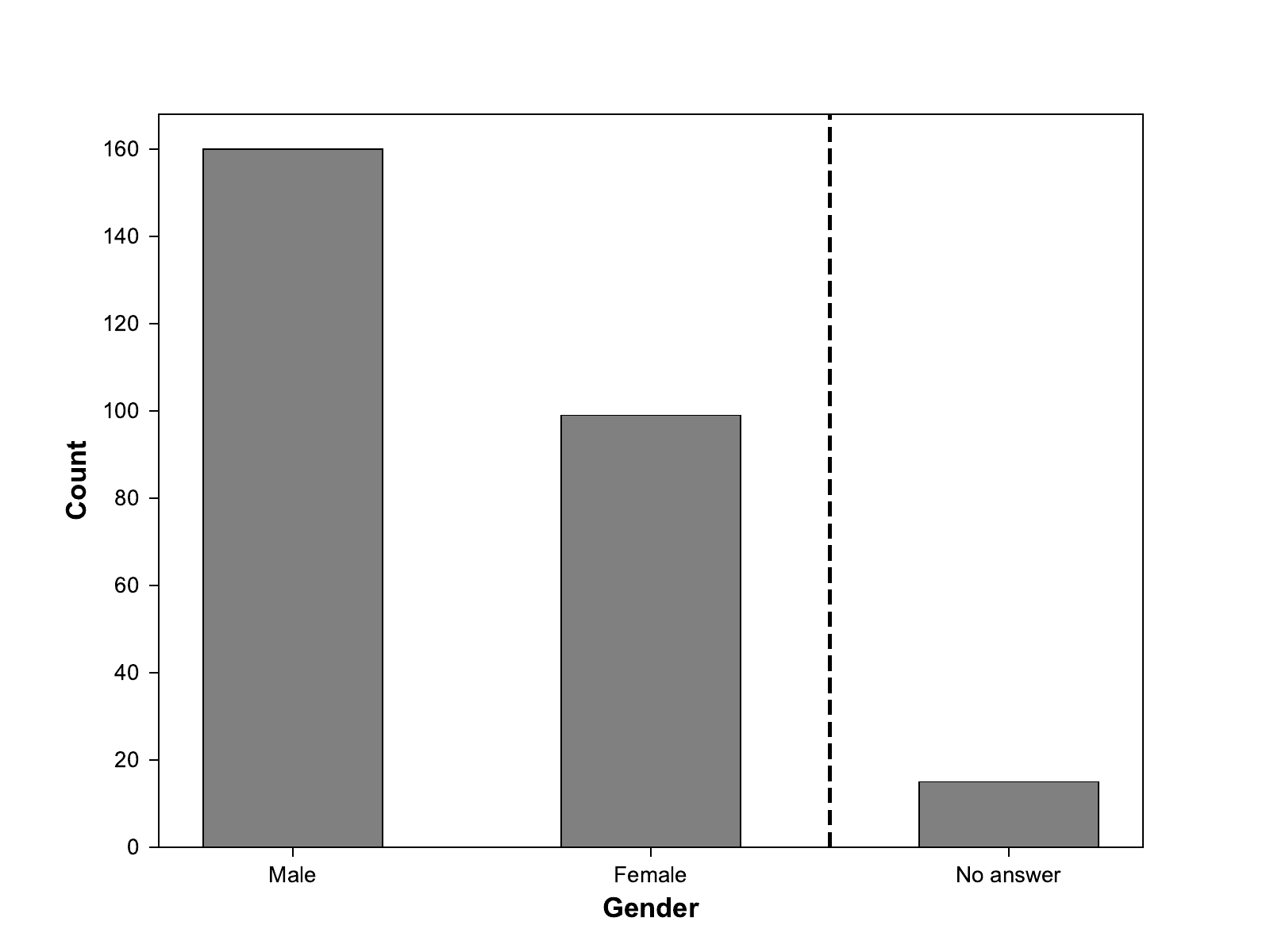}
  \caption{Gender of respondents}
  \label{fig:gender}
\end{figure}

Figure \ref{fig:gender} shows that most of the respondents were male (62\%). In terms of education, a majority of participants were educated to at least degree level, with more than 67\% having at least a Bachelor's degree and nearly 8\% of all respondents held a doctorate, again this is likely to be influenced by the researchers' personal and professional networks. Our sample also represented a wide range of employment sectors. These include Accounting/Finance, Administration, Engineering, Human Resources, Education, Legal and Sales/Marketing. There is a fairly even distribution among all of the sectors (ranging from 2-7\%), with the exception of IT/Computing, which contains a clear majority (27\%).

\begin{table*}[ht!]
\centering
\begin{tabular}{lllll}
\hline
\multicolumn{3}{|l|}{\textbf{}}                                                                                                    & \multicolumn{2}{c|}{\textbf{Comfortable held by:}}                                                                                                                                        \\ \hline
\multicolumn{1}{|l|}{\textbf{Biometric}}   & \multicolumn{1}{l|}{\textbf{Heard of}} & \multicolumn{1}{l|}{\textbf{Knowingly used}} & \multicolumn{1}{l|}{\textbf{\begin{tabular}[c]{@{}l@{}}Government \\ department\end{tabular}}} & \multicolumn{1}{l|}{\textbf{\begin{tabular}[c]{@{}l@{}}Private \\ company\end{tabular}}} \\ \hline
\multicolumn{1}{|l|}{\textbf{Fingerprint}} & \multicolumn{1}{l|}{211}               & \multicolumn{1}{l|}{146}                     & \multicolumn{1}{l|}{72}                                                                        & \multicolumn{1}{l|}{125}                                                                 \\ \hline
\multicolumn{1}{|l|}{\textbf{Palm print}}  & \multicolumn{1}{l|}{136}               & \multicolumn{1}{l|}{20}                      & \multicolumn{1}{l|}{39}                                                                        & \multicolumn{1}{l|}{69}                                                                  \\ \hline
\multicolumn{1}{|l|}{\textbf{Hand vein}}   & \multicolumn{1}{l|}{31}                & \multicolumn{1}{l|}{0}                       & \multicolumn{1}{l|}{21}                                                                        & \multicolumn{1}{l|}{36}                                                                  \\ \hline
\multicolumn{1}{|l|}{\textbf{Face}}        & \multicolumn{1}{l|}{179}               & \multicolumn{1}{l|}{88}                      & \multicolumn{1}{l|}{54}                                                                        & \multicolumn{1}{l|}{91}                                                                  \\ \hline
\multicolumn{1}{|l|}{\textbf{Retina}}      & \multicolumn{1}{l|}{196}               & \multicolumn{1}{l|}{38}                      & \multicolumn{1}{l|}{31}                                                                        & \multicolumn{1}{l|}{77}                                                                  \\ \hline
\multicolumn{1}{|l|}{\textbf{Iris}}        & \multicolumn{1}{l|}{160}               & \multicolumn{1}{l|}{24}                      & \multicolumn{1}{l|}{32}                                                                        & \multicolumn{1}{l|}{62}                                                                  \\ \hline
\multicolumn{1}{|l|}{\textbf{Signature}}   & \multicolumn{1}{l|}{200}               & \multicolumn{1}{l|}{158}                     & \multicolumn{1}{l|}{129}                                                                       & \multicolumn{1}{l|}{133}                                                                 \\ \hline
\multicolumn{1}{|l|}{\textbf{Gait}}        & \multicolumn{1}{l|}{82}                & \multicolumn{1}{l|}{4}                       & \multicolumn{1}{l|}{19}                                                                        & \multicolumn{1}{l|}{33}                                                                  \\ \hline
\multicolumn{1}{|l|}{\textbf{Typing}}      & \multicolumn{1}{l|}{87}                & \multicolumn{1}{l|}{22}                      & \multicolumn{1}{l|}{31}                                                                        & \multicolumn{1}{l|}{44}                                                                  \\ \hline
\multicolumn{1}{|l|}{\textbf{Voice}}       & \multicolumn{1}{l|}{189}               & \multicolumn{1}{l|}{65}                      & \multicolumn{1}{l|}{59}                                                                        & \multicolumn{1}{l|}{66}                                                                  \\ \hline
                                           &                                        &                                              &                                                                                                &                                                                                         
\end{tabular}
\caption{The breakdown of which biometric methods a participant had heard of, knowingly used and would be comfortable being held by a government department or a private company.}
\label{tab:heard_used_held}
\end{table*}

\begin{figure}[ht!]
\centering
   \includegraphics[width=0.85\linewidth]{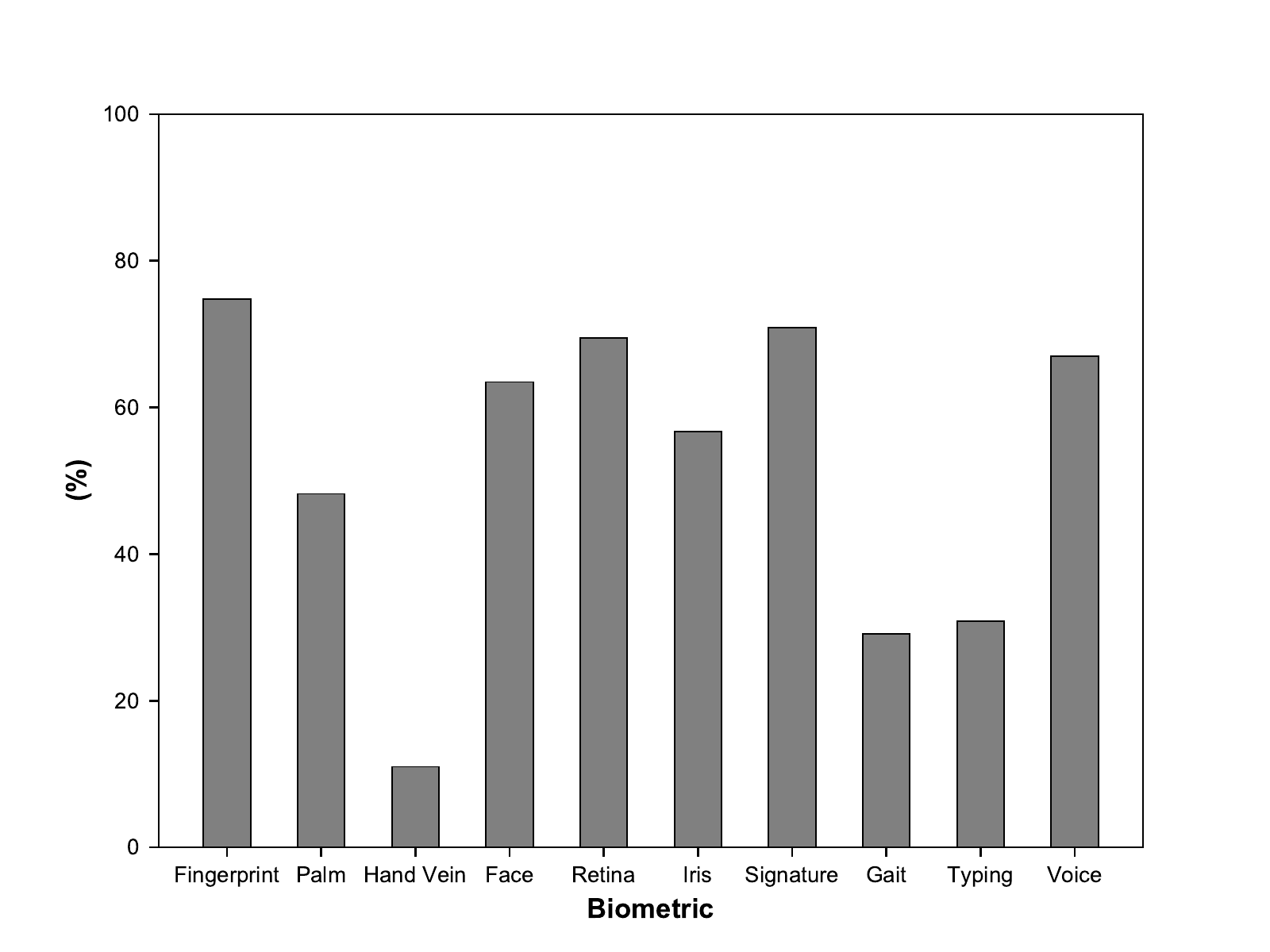}
   \caption{Have you heard of this biometric?}
   \label{fig:heard_of} 
\end{figure}

\begin{figure}[ht!]
\centering
   \includegraphics[width=0.85\linewidth]{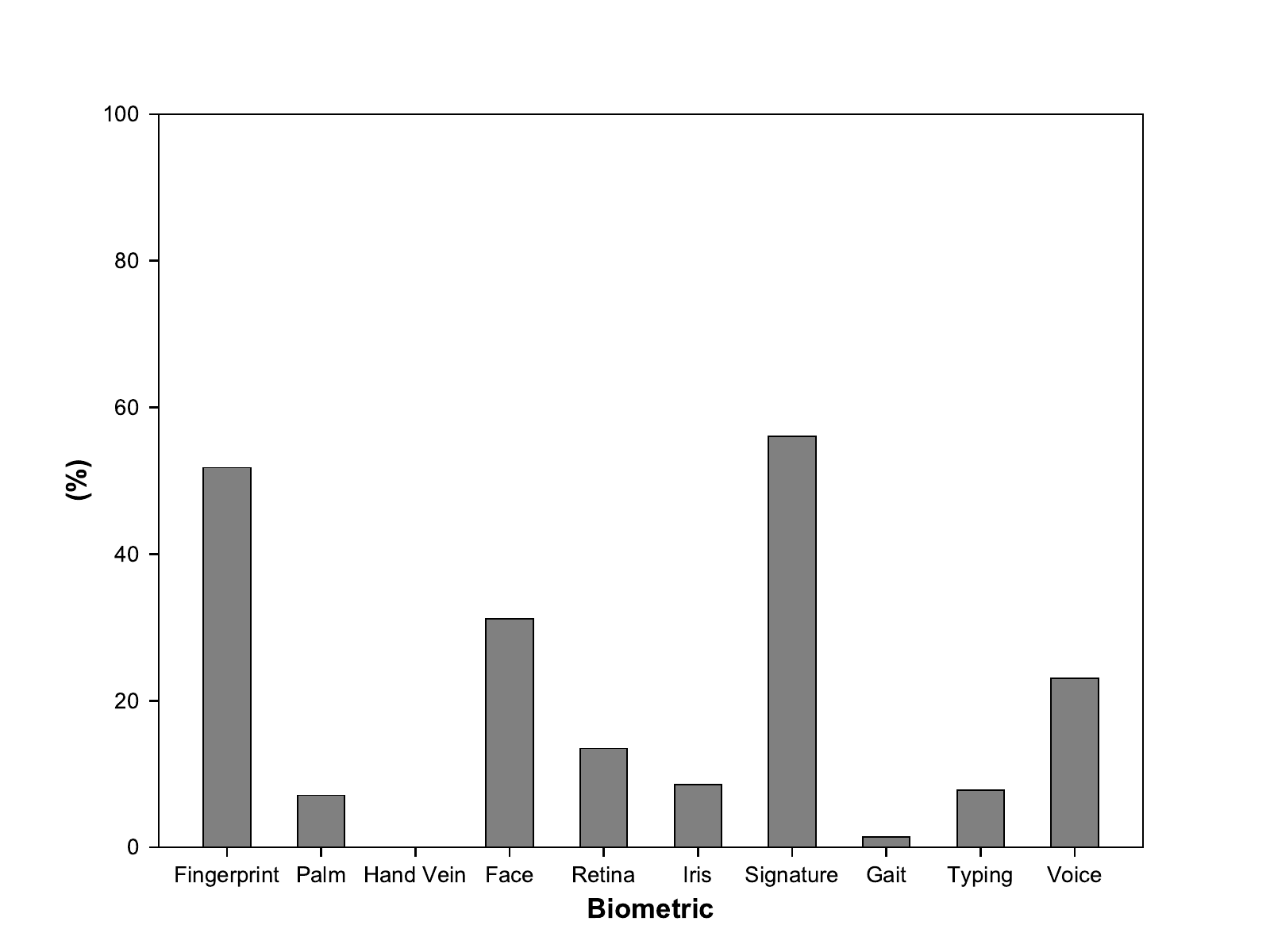}
   \caption{Have you ever knowingly used this biometric?}
   \label{fig:knowingly_used}
\end{figure}

\begin{figure}[ht!]
\centering
   \includegraphics[width=0.85\linewidth]{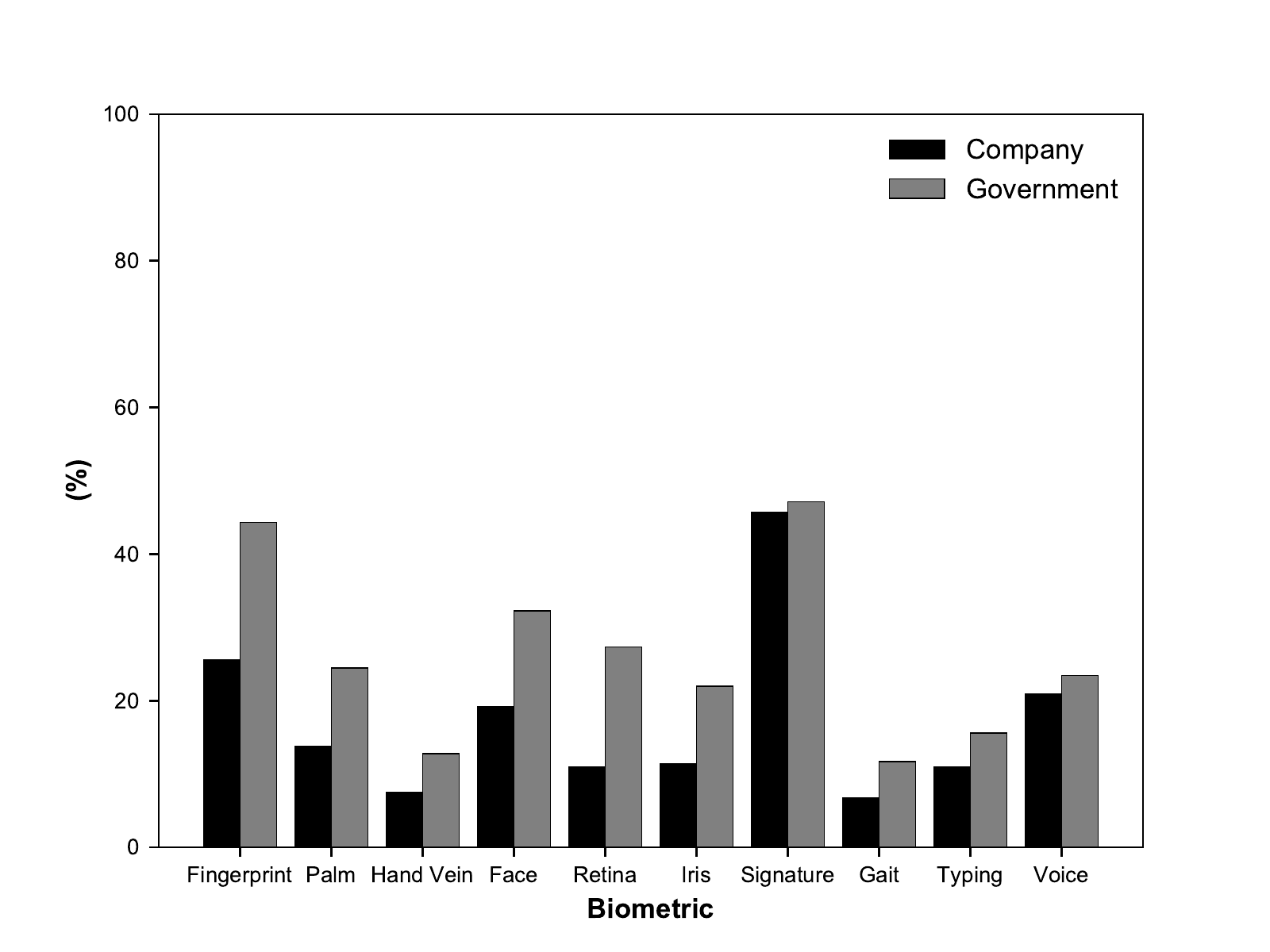}
   \caption{Would you be comfortable with a company or government holding this data?}
   \label{fig:company_or_government}
\end{figure}

One of the first areas we examined was participant's knowledge of biometric systems. We presented participants with a list of common biometrics (both physical and behavioural) and asked them to indicate if they had heard of the scheme (Figure \ref{fig:heard_of} and Table \ref{tab:heard_used_held}), whether they had knowingly used it (Figure \ref{fig:knowingly_used}) and whether they would be comfortable with it being held by a company or government (Figure \ref{fig:company_or_government}).

The first thing to note is that broadly speaking the participants had a good knowledge of a number of physical biometric technologies, as seen in Figure \ref{fig:heard_of}. The most commonly known of technology is the use of fingerprints, which is unsurprising given their prevalence in personal devices and our daily lives (e.g., smartphones or immigration at an airport). 

At the other end of the scale, those that fewer participants had heard of, were gait, typing and hand vein. It is interesting to note that two of these methods (gait and typing) are arguably classified as behavioural biometrics, which perhaps suggests that there is a lack of awareness around these approaches. \fixmetext{This is potentially unsurprising in that traditionally behavioural biometrics do not require the user to interact with any specific hardware directly. Instead their behaviours are normally monitored remotely.} This is further underlined by comparing the results to work of Furnell and Evangelatos \cite{furnell2007}, where keystroke dynamics also proved to be amongst the biometrics of which the fewest participants were aware. 

\begin{table*}[ht!]
	\centering
    \begin{tabular}{|p{2cm}|p{2cm}|p{2cm}|l|p{2cm}|l|}
    \hline
    ~               & \textbf{Banking} & \textbf{Online Shopping} & \textbf{Airport} & \textbf{Mobile Device} & \textbf{Home} \\ \hline
    \textbf{1}               & 85      & 2               & 81      & 16            & 18   \\ \hline
    \textbf{2}               & 90      & 28              & 42      & 21            & 21   \\ \hline
    \textbf{3}               & 25      & 69              & 36      & 39            & 33   \\ \hline
    \textbf{4}               & 2       & 54              & 24      & 83            & 39   \\ \hline
    \textbf{5}               & 0       & 49              & 19      & 43            & 91   \\ \hline
    \textbf{Average Rank} & \textbf{1.72}    & \textbf{3.59}            & \textbf{2.29}    & \textbf{3.57}          & \textbf{3.81} \\ \hline
    \end{tabular}
    \caption{The breakdown of rankings for each situation. Each cell shows the total number of participants who had ranked the situation with that specific need for security (1 to 5). A total of 202 participants answered this question.}
    \label{tab:rankings}
\end{table*}

The results in Figure \ref{fig:knowingly_used} and Table \ref{tab:heard_used_held} highlight the methods that have been knowingly used by participants. Some of the most commonly used are fingerprint and facial recognition. Both of these technologies have been used to secure personal devices and so such a common usage is expected. However, the most surprising result is that the most commonly used biometric was `signature'. It is highly likely that many participants have mistaken a simple signature (e.g., signing for a parcel) with the biometric method. This is reinforced by Furnell and Evangelatos \cite{furnell2007} who identified a similar case of potential confusion. Very few participants claimed to have knowingly used typing, voice and gait as biometrics. This is quite an intriguing point to draw out as these are indeed all biometrics that can be recorded remotely, so it is entirely possible (and likely in some cases) that these have been used by the participants without them actually having realised it. As an example, a number of banks are using voice recognition as part of their online banking services (e.g., Barclays Bank \cite{bbcBarclaysVoice2016} and HSBC~\cite{tele17} in the UK). 

Figure \ref{fig:company_or_government} (and Table \ref{tab:heard_used_held}) provides a comparison between the data that participants would be comfortable being stored by a government or a private company. From the results, we can see that respondents were uniformly more comfortable with their biometric data being held by a government, rather than a private company. This is a particularly interesting when considering fingerprints, which although potentially  held on a variety of governmental databases, it is highly likely that a significant number of participants already trust their fingerprint data to a number of a device providers (e.g., Apple, Google or Samsung). This in itself raises other concerns, particularly in the future when we consider users understanding privacy implications of widespread adoption of biometric technologies \cite{williams2016future}.   

The next question looked to understand which biometrics were considered to be the most secure. In the first instance participants were asked to rank five situations in order of their need for security, which allowed a ranking of the scenarios to be calculated drawing out those situations that were perceived to have the greatest need for security. An average ranking was calculated for each of the different situations; we note here that 202 out of the 282 participants chose to answer this question. 

First the number of participants that had ranked each situation at any given level was calculated. This gave a total ranking for each of the situations, where the lower number dictates the situation with the perceived greatest need for security. This was then used to calculate an average ranking for each situation, and determine the situation that was deemed to have the greatest need for security and the situation with the least need for security. The breakdown of the resulting rankings of each situation can be seen in Table \ref{tab:rankings}, with the overall rankings as follows: (1) Banking; (2) Airport; (3) Online shopping; (4) Mobile device; and (5) Home.

The results of this ranking are intriguing for several reasons. It is perhaps not surprising that banking was ranked as the situation with the greatest need for security. However, the fact that home was ranked as the least need for security was particularly surprising as our own home is the place where we are meant to feel the most safe and secure. Further investigation revealed that approximately 58\% of respondents actually ranked their mobile device as requiring a greater level of security than their home. This is a particularly insightful discovery and highlights just how essential mobile devices have become in our everyday lives. The rankings of each of the situations was found to be consistent across all of the age groups, with no real variation across different ages, suggesting that these are universal opinions (at least within our sample).

Following this, participants were asked which biometric technology that they would be comfortable using for each of the particular situations from the previous question. In order to determine which technology was perceived to be the most secure, amongst the respondents, analysis was carried out to understand the biometric that each participant perceived to be the most appropriate for the five situations, shown previously. For example, if a participant had ranked `\textit{airport}' as the greatest need for security and then selected facial recognition as the appropriate technology for this situation, then this would be inferred to be the technology that the participant considered to be the most secure. 

\begin{figure}[ht!]
\centering
   \includegraphics[width=0.85\linewidth]{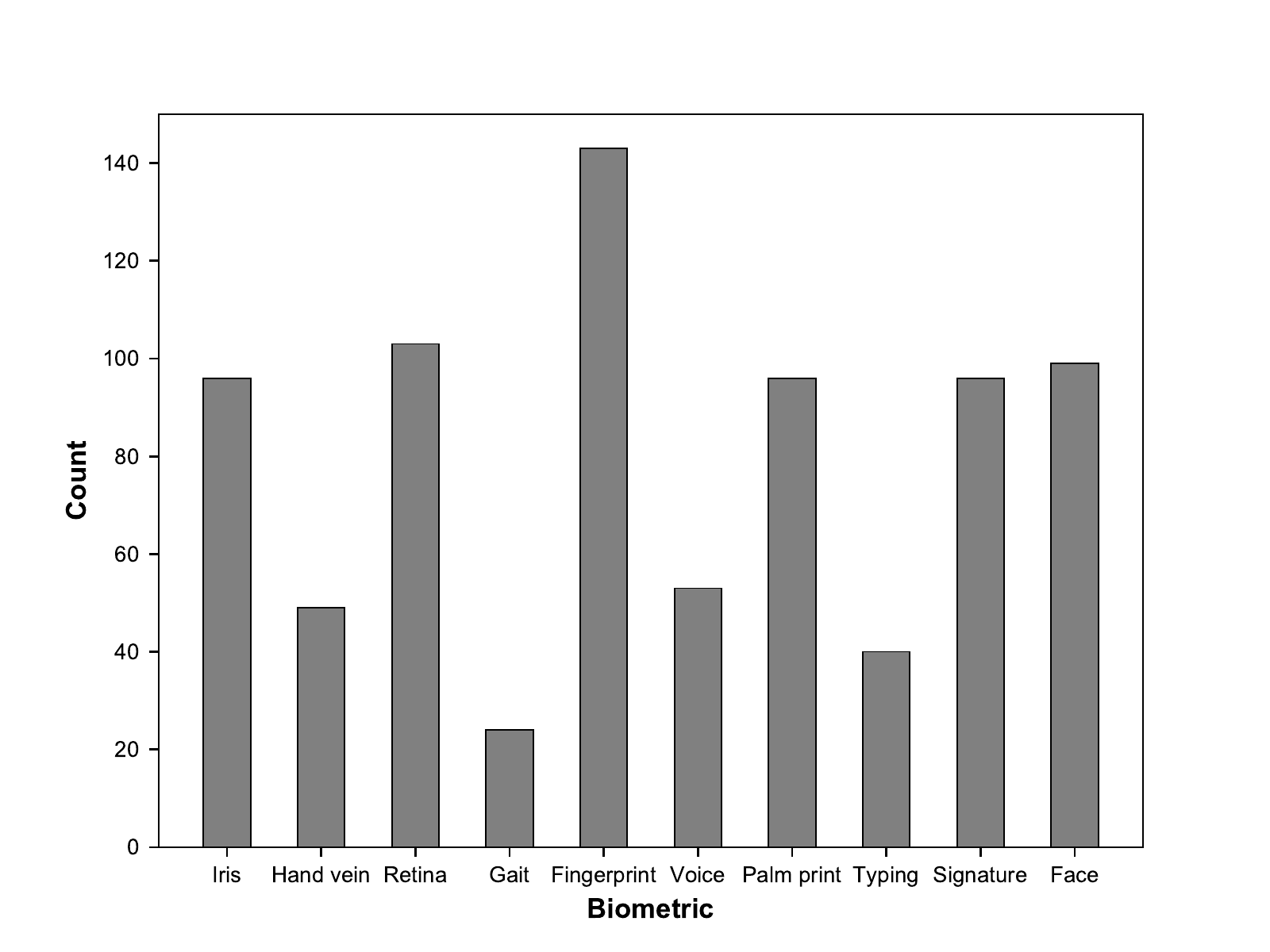}
   \caption{The biometrics that were ranked as the most secure}
   \label{fig:top_ranked} 
\end{figure}

\begin{figure}[ht!]
\centering
   \includegraphics[width=0.85\linewidth]{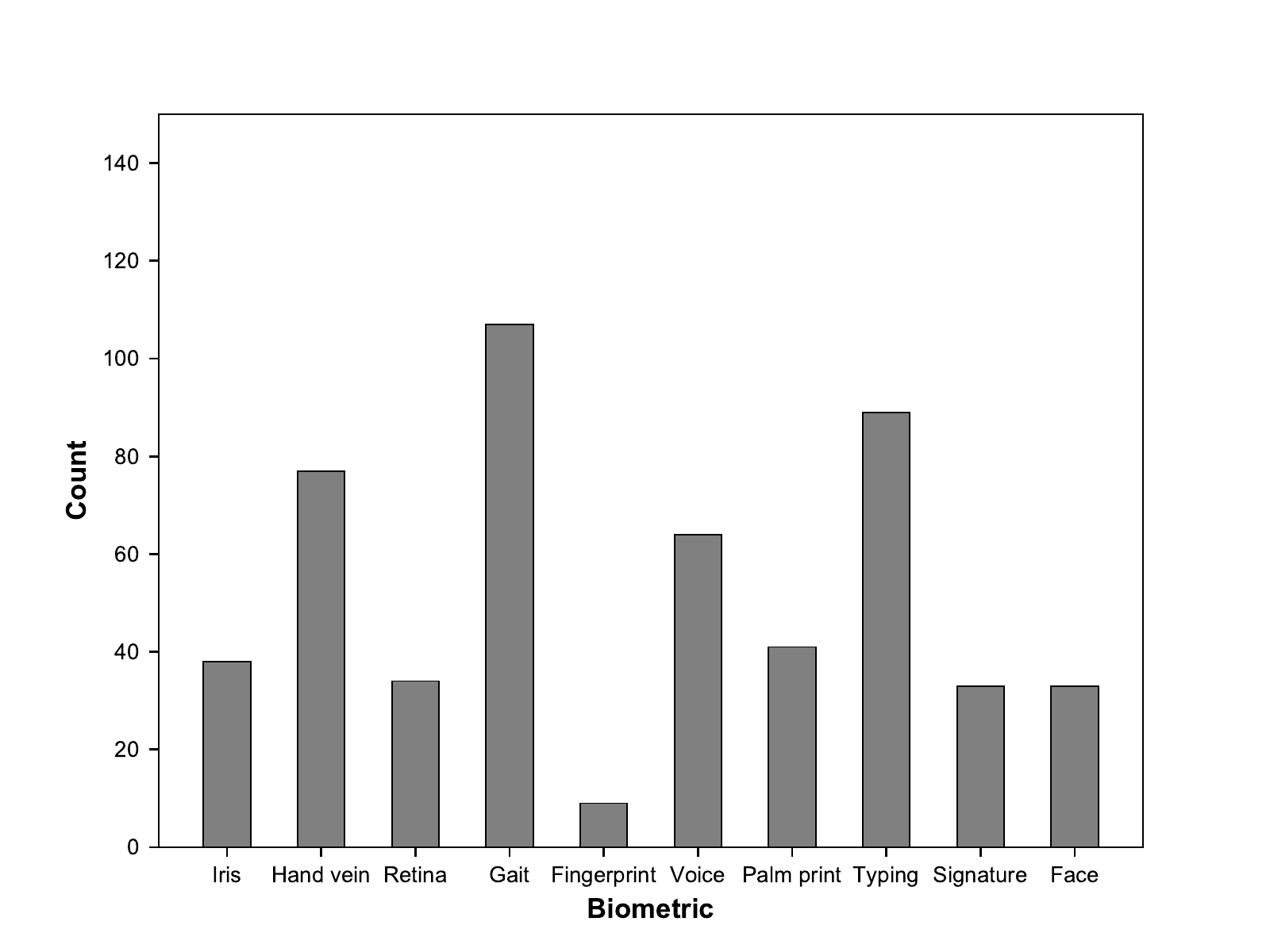}
   \caption{The biometrics that were ranked as the least secure}
   \label{fig:bottom_ranked}
\end{figure}

The results of this analysis can be seen in Figures \ref{fig:top_ranked} and \ref{fig:bottom_ranked}, which shows distribution of biometrics selected as suitable for the most secure situation (Figure \ref{fig:top_ranked}) and those that were selected as not suitable for any of the situations (Figure \ref{fig:bottom_ranked}). 

Firstly, focusing on those that were ranked as the most secure biometrics, it becomes apparent that there is a trend towards those methods that the participants were most familiar with. For example, the method most clearly perceived as the most secure was fingerprint recognition. This is something that is increasingly common in our daily lives, for example, the large majority of personal devices (e.g., smartphones and tablets) now make use of fingerprint recognition and this is now becoming a common feature of international travel. Fingerprint recognition was also the biometric that the majority of participants had heard of and knowingly used, as seen in Figure \ref{fig:heard_of}, Figure \ref{fig:knowingly_used} and Table \ref{tab:heard_used_held}. However, these fingerprint sensors in our personal devices are easily compromised, as demonstrated by both Yang et al. \cite{yang2016iphonevulnerability} and Kanchikere and Sudha \cite{kanchikere2018hacking}. This highlights that a lack of awareness of other biometric methods is potentially damaging to the personal security of users. 

Another observation that can be made is that the biometrics that were perceived to be the most secure (iris, retina, palm print and facial recognition) are those methods that are perhaps either most commonly used in our day-to-day lives or those that are most common in popular culture. For instance, it is very common to see palm print recognition or retinal scans on our screens. However, in the case of facial recognition, as with fingerprint recognition, previous iterations of this technology has been shown to be easily compromised. For example, facial recognition has previously been circumvented with the use of printed masks/photos \cite{galbally2016three} \cite{biometricks17}. 

At the other end of the scale it was noteworthy that the two behavioural biometrics (gait and typing analysis) were considered to be amongst the least secure. In addition to the two behavioural biometrics, voice recognition was also considered to be amongst the least secure. As previously discussed, three of these methods were all methods that had not been `knowingly used' by the majority of participants and are all methods that could be used remotely without the user's knowledge. It is entirely possible that the intangible nature of these methods contributes to their perceived lack of security. Finally, hand vein recognition is amongst those method perceived to be the least secure. This is perhaps unsurprising given that this was the biometric that the fewest number of respondents had actually heard of, with none of the participants having knowingly used this method.

The final section of the survey asked participants for their opinions on the security of biometrics when compared to other authentication methods. First participants were asked whether they thought biometrics were as secure as passwords with 83\% of participants agreeing that this was the case. Participants were then asked whether they thought that biometrics could provide the same level of security as two-factor authentication, with approximately 75\% of people agreeing with this statement. Finally, participants were asked whether they felt that biometrics could be easily compromised. Only 46\% of participants believed that biometrics were not easily compromised, which was a surprising result. This is especially true when considering that the majority of respondents felt that biometrics were as secure of two-factor authentication or passwords. 

One of the key discoveries of this survey was that the participants generally felt that the methods they were most familiar with (e.g. fingerprints) were the most secure. This is perhaps not surprising but does highlight that familiarity exposure to these technologies helps to generate support for the methods. 
    
\subsection{Thematic and Word2Vec Analyses}
\label{subsec:word2vec_analysis}
To complement the research in Section~\ref{subsec:survey_results}, we were also keen to examine people's understanding of the meaning of the term \textit{biometrics} itself. In our survey, we had asked participants whether they knew what was meant by the term \textit{biometrics}, with 74\% of respondents claiming to know what the term meant. Following this question these participants were asked to provide a definition of what they thought was meant by the term; this was completed by 49\% of participants overall. This section analyses the responses that were given using two analysis techniques. First, we apply thematic analysis, which enables key themes to be discovered in the data \cite{braun2006using}. After this, we explore the utility of an automated analysis approach called Word2Vec \cite{mikolov2013efficient}; this technique models words in a vector space to allow for insight into textual content. Finally, we compare the findings of each technique to published definitions of biometrics today (e.g., \cite{bolle2013guide} \cite{oed} \cite{isobiom}), which have been analysed using thematic analysis.

Definitions provide a simple yet effective method to elicit an individual’s understanding of a particular term. We received a variety of definitions from participants for biometrics, which after analysis, resulted in several key themes. The most prominent of theme was that of \textit{identity}, with biometrics being viewed as a means to identify, or verify the identity of, an individual. As one participant stated, it is the ``use of biological/physical parameters to identify a person''. This therefore highlights a primary application of biometrics today according to individuals. 

The second most significant theme also features in the quote above, i.e., the central part played by \textit{biological and physical characteristics} within the identification process. According to participants, identification is thus not based on what a person knows (e.g., with passwords), it is driven by who they are, or in the words of another participant ``[biometrics covers] essentially data about who you are/your body''. These types of characteristics and data, along with the examples provided by participants (e.g., fingerprints, retina scans), match many of the types presented later in the study. Several participants even used examples of biological data as their definition for biometrics. While the general themes of identity and human characteristics were consistent, one participant made a point that biometrics allowed for identification ``through non-traditional means''. This view, although isolated, hints to a perception of biometrics as not yet fully streamlined as the new standard in identification.

Biometrics were also viewed as \textit{a measure of} (or, statistics based on) key characteristics within the biological and physical space. To quote one participant, ``a measure of the body using certain characteristics such as DNA or gait of walk that are individual to each person''. This also highlights a \textit{uniqueness} component, which enables biometric features to be somewhat distinguishing; a primary factor motivating their use in identification. Other noteworthy themes which emerged---albeit featuring significantly less---included \textit{biometrics as behaviour}, and \textit{biometrics for the purposes of security}. In the former theme, participants expressed that the behavioural characteristics of individuals were also central to the understanding of biometrics and how they are used. In most cases where behaviour was mentioned, it was alongside physical characteristics, for instance, biometrics are ``related to people's physical and behavioural characteristics such as fingerprints, retinal scan, movement''.  In this case, physical characteristics were fingerprints and retinal scan, while the behavioural characteristic was movement (e.g., gait). 

For the latter theme, biometrics were closely associated with security. As stated by a participant, biometrics were a means to ``identify an individual by his unique biological identity / characteristic for security purposes''. This pulls together several of the previous themes including identity, biological characteristics and uniqueness. One participant extended the role of security in biometrics further by describing it as ``security measures based upon the human body''. With another, it was perceived as ``the use of body characteristics as an extra `factor' in authentication''. These views provide additional insight into how some individuals regard biometrics and topics including security, authentication and authorisation (mentioned in another quote).

After analysing the content manually using thematic analysis, we were interested in exploring the extent to which automated methods may be able to extract similar, or more nuanced, themes. This could also verify the findings of the manual analysis, or indeed, increase the confidence that may be placed in automated techniques in future analyses. 

For this task, we decided to analyse the definitions using Word2Vec \cite{mikolov2013efficient}, an approach that models words in a vector space. This method takes a large corpus of text as its input, in this case the model was created using the Google News data (containing around 100 billion words), which includes 300-dimensional vectors for 3 million words and phrases \cite{mikolov2013distributed}. This input is used to develop word vectors; first stop words (e.g., and, the, but) are removed, a vocabulary is constructed from the training text and then vector representations are identified for each word. Words can then be positioned within the vector space. Those words that are in close proximity to one another within the vector space share a common semantic meaning or context. A key point to note about this approach is that it derives context and relationships within the text based on the training corpus. As the corpus is created from publicly available text, it is free from biased on opinions and experiences of the researchers.

\begin{figure}[ht!]
	\centering
	\includegraphics[width=1.1\linewidth]{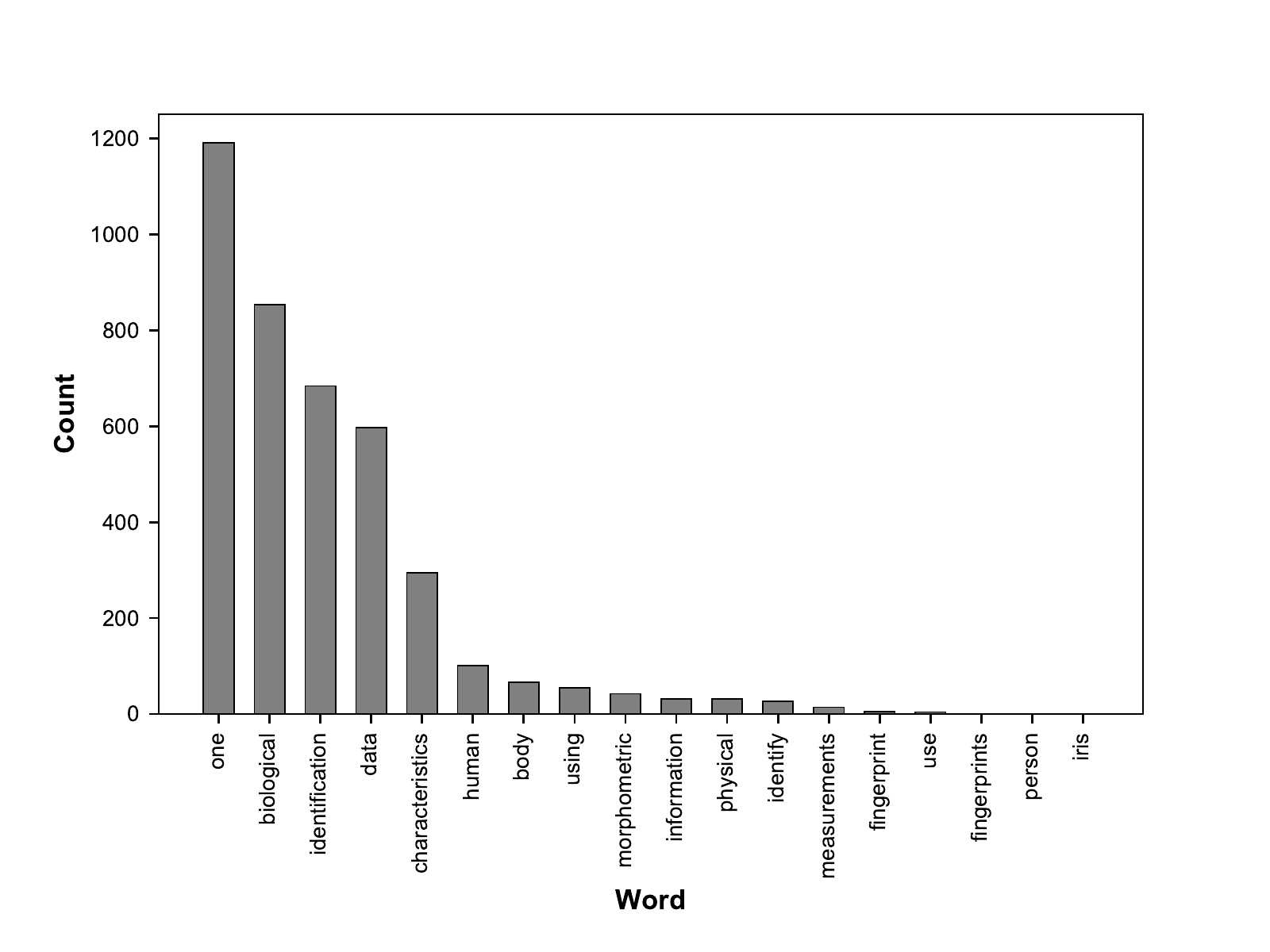}
	\caption{A graph of the most common word centroids based on the clustering approach}\label{fig:centroids}
\end{figure}

For our use of the method, we first took the biometrics definition from each individual and tokenised it into individual words, with stop words removed. Each of the remaining words was then projected onto vector space. The resulting word vectors were clustered, using $k$-means clustering \cite{hartigan1979kmeans} to determine the key concepts involved in the public perceptions of biometrics. It was determined, using the gap statistic approach \cite{tibshirani2001}, that the optimal grouping of the data would be into four clusters. This implied that the provided definitions could be distilled down into four key concepts. As $k$-means clustering is a stochastic approach it was repeated 1000 times. This resulted in 4 centroids in vector space per iteration. Each centroid can then be mapped back to find the work closest to its vector representation. These representations are shown in Figure \ref{fig:centroids} for the 1000 iterations. 

As can be seen in Figure \ref{fig:centroids}, the five most common themes/terms that can be drawn out from the provided definitions are: one, biological, identification, data and characteristics. The most popular term, \textit{one}, can be understood to refer to oneself, or the individual. This could be taken to refer to the subject from whom biometrics may be attained. In the second and third terms, we can see the \textit{biological} nature of biometrics being emphasised as well as their utility for \textit{identification} purposes. The importance of \textit{data} and identity \textit{characteristics} to biometrics is highlighted in the two remaining prominent terms. These complement the previous themes and together allow for a clearer depiction of how participants understand and perceive modern-day biometrics. 

To compare these findings with those from the manual analysis, there are several similarities. Most notably, the themes of identification, and biological and physical characteristics are prominent in both analyses. The starkest difference from a superficial perspective is the prominence of the term `one' in the Word2Vec analysis. As discussed above however, this can be interpreted as relating to an individual i.e., the subject of the biometric. This would therefore align with the findings of the thematic analysis. 

A less compensable difference in the two approaches is the importance of measures (as a theme) in the manual analysis, but not as high a ranking of measurements/metrics in the Word2Vec's clusters. This could be due to a significant number of unknown or new words. Word vector analysis, such as Word2Vec is known to struggle with words that are out of vocabulary (OOV), if the model has not encountered the word before then it will not know how build a vector representation. With this specific dataset there is a range of specialist vocabulary in use and as such may be OOV. Similarly, the accuracy of the vectorization process can be impacted if the Word2Vec model contains no shared representations at sub-word level. Word2Vec represents every word as an independent vector, despite many words being morphologically similar. 

The next step in our research was to reflect on the perceptions and understanding of biometrics by our participants, as compared to the meaning of the term supplied by official and well-publicised sources. This would enable us to gauge how accurate---or rather, aligned---participants' perceptions were in light of standard definitions. For this task we used 16 sources: the four most well-known English dictionaries \cite{oed}\cite{camup}\cite{marweb}\cite{collinsdict}; four standard-setting organisations \cite{isobiom}\cite{nist09}\cite{itubiom}\cite{bsibiom}; three governmental organisations and a biometrics institute \cite{hobio}\cite{dhsbio}\cite{gdprbio}\cite{biobio}; and four prominent texts \cite{bolle2013guide}\cite{anderson2008security}\cite{coats2007practitioner}\cite{jain2007handbook}. These specific sources were selected because of either their nature, pedigree, or general need to be accessible to individuals of the public. After collecting definitions from each source, we applied thematic analysis (identical to our manual analysis of participants' data) to extract the core themes in the dataset. 

From the completed analysis, we found a notable overlap in the themes arising from participant definitions and the official definitions. The most common themes in the official definitions were individuals or people who the biometrics would relate to, behavioural, biological and physical characteristics, and biometrics' use for identification and recognition. Merriam-Webster captures these themes aptly in defining biometrics as ``the measurement and analysis of unique physical or behavioral characteristics (such as fingerprint or voice patterns) especially as a means of verifying personal identity'' \cite{marweb}. This description itself also highlights many of the themes that were identified in our participants' responses. As such, we might conclude that participants generally perceive and understand biometrics accurately, or at least that their understanding aligns with popular conceptualisations. There was one area where there was some disparity, however, i.e., behavioural characteristics and their importance to biometrics. In the official definitions, we found that behaviour featured in a majority of sources. If we consider participant responses however, behaviour was rarely discussed and there was a clear emphasis on biological or physical characteristics. This point further supports our survey results (as well as prior work by Furnell and Evangelatos \cite{furnell2007}) regarding the lack of widespread awareness of behavioural biometrics. 

This analysis of the individual definitions of the term \textit{biometrics} has drawn out a number of interesting concepts and themes. It is particularly interesting to note that it would appear, as a result of this analysis, that the general public have a good grasp of most of the key areas surrounding the concept of biometrics. This with the exception of the behavioural component which we have discussed above.

\section{Conclusions and Future Work}
\label{conclusions_and_future_work}
The survey has highlighted that there is seemingly a good level of awareness and acceptance of certain biometric methods. The participants had heard of the vast majority of methods, with only hand vein recognition being under represented, although this is perhaps understandable given that this technology is not particularly widespread. While there are significant differences between the underlying technologies of hand vein and palm print recognition the method of collection will look similar to the average person. 

It is notable that while there is a general acceptance of these technologies, it is very much dependent on the context. However, the research has highlighted that users are seemingly the most comfortable with those methods that are more commonplace and familiar (e.g., fingerprint or facial recognition). When considering methods that are slightly more intangible (e.g., typing or gait analysis) they are typically less well regarded or understood. This suggests that perhaps there is scope to further develop the public's acceptance of these methods. 

From our manual analysis of the definitions of biometrics provided by participants, we identified several key themes core to people's understanding of the topic. The complementary automated approach, Word2Vec analysis, was also shown to be able to highlight central themes in the definition data. This is commendable and could increase confidence in the approach's further use. Another key finding of our study is that, largely speaking, the definitions provided by participants closely aligned to the core elements of the more official definitions of biometrics. The exception to this being the behavioural component, which was also a result from our survey data.

Future work in this area would look to further develop a deeper understanding of the experiences and perceptions of biometrics technologies amongst the general public. This could be achieved by larger more representative samples, and cover wider ranges of biometrics and related technologies \fixmetext{which we were unable to address in this article; for instance, it could be interesting to investigate EEG and cardiac biometrics.} Additionally, one of the key areas for future work is to establish methods for improving acceptance and understanding of the less tangible, behavioural biometric methods.


\bibliographystyle{model1-num-names}
\bibliography{sample.bib}







\end{document}